\documentstyle[aps,prl]{revtex}
\begin{document}
\input epsf.sty
\twocolumn[\hsize\textwidth\columnwidth\hsize\csname %
@twocolumnfalse\endcsname
\draft
\widetext

\title{Spin-Peierls Transition in  CuGeO$_3$:  Critical,
Tricritical or Mean Field? }
\author{ R. J. Birgeneau, V. Kiryukhin, and Y. J. Wang} 
\address{Department of Physics, and
Center for Materials Science and Engineering,
Massachusetts Institute
of Technology, Cambridge, MA 02139}
\date{\today}
\maketitle

\begin{abstract}  

The spin-Peierls phase transition in CuGeO$_3$  has
been extensively studied utilizing a variety of experimental
techniques.  Interpretations of the phase transition behavior vary
from tricritical to mean field to Ising critical to XY critical.  We
show that the behavior in the vicinity of the phase transition
of each of the order parameter, the magnetic energy gap and the
heat capacity can be  quantitatively fitted with few
adjustable parameters with a mean field model incorporating a
tricritical to mean field critical crossover in the transition
region.
\end{abstract}

\phantom{.}
]
\narrowtext

\section{Introduction}

The spin-Peierls transition corresponds to the dimerization of a
one-dimensional S = $1\over 2$ antiferromagnetic chain coupled to a
three dimensional elastic medium \cite{SP}-\cite{Mon}.  
Until relatively recently,
spin-Peierls transitions had only been observed in organic charge
transfer compounds such as copper bisdithiolene (TTF-CuBDT) 
\cite{Jac}-\cite{Jac1}.
Experimental information obtainable in such systems has been limited
both by the size of available single crystals and by the sensitivity
of these materials to damage by x-rays or electrons.  Nevertheless,
some important information on the spin-Peierls phase transition has
been obtained in a number of different organic materials.
Interestingly, in most, if not all cases, the data are consistent with
a simple BCS-type mean field transition \cite{Jac}-\cite{Jac1}.

Much more complete experimental work on the spin-Peierls transition
has been made possible by the discovery that a structurally simple,
inorganic chain compound copper germanate (CuGeO$_3$) undergoes a
spin-Peierls transition  at a transition temperature around 14K \cite{Hase}.
The crystal structure of CuGeO$_3$ is orthorhombic, space group {\it
Pbmm}, with a unit cell of dimensions $a = 4.81$ \AA, $b = 8.47$ \AA
\  and $c = 2.94$ \AA \  at room temperature \cite{Voll}.  
The Cu$^{2+}$ ion carries
a spin $S = {1\over 2}$ and forms a (CuO$_2$) chain with the neighboring
Cu$^{2+}$
ions along the $c$-axis direction.  The successive Cu$^{2+}$ S =
$1\over 2$ spins are antiferromagnetically coupled through the
superexchange interactions via the bridging oxygen atoms.  Below the
spin-Peierls transition temperature, $T_{SP}$, the dimerization of
Cu-Cu pairs along the $c$-axis direction, accompanied by shifts of the
bridging oxygen atoms in the $ab$ plane, gives rise to superlattice
reflections at the $({h\over 2},\ k,\ {l\over 2})$ ($h, l$: odd and $k$:
integer) reciprocal-lattice positions \cite{Hirota}.  These have been
observed in electron diffraction \cite{Kam}, x-ray \cite{Pouget}, and elastic
neutron scattering \cite{Hirota} experiments.  Using coarse resolution
x-ray diffraction techniques, Pouget {\it et al.} \cite{Pouget} have
measured the pretransitional thermal lattice fluctuations
whose correlation lengths diverge anisotropically
with decreasing temperature in a manner consistent with mean
field theory.  These same fluctuations have been studied at
high resolution using synchrotron x-ray diffraction techniques
by Harris {\it et al.} \cite{Harris}. These latter authors observe within
about 1K of
$T_{SP}$ large length scale fluctuations with characteristic length
scales about an order of magnitude longer than those characterizing
the bulk critical fluctuations.

In spite of this large amount of work, it is still not agreed whether
the observed transition behavior reflects mean field or
critical behavior.  Extant models include: a) tricritical to 3D Ising
crossover behavior \cite{Harris,Werner}; 
b) mean field behavior \cite{Werner}; c) 3D XY with
corrections to scaling \cite{Lum}, and, most exotically, d) a 2D XY to 3D
XY crossover as
$T_{SP}$ is approached \cite{Lorenzo}. 
Harris {\it et al.} \cite{Harris} first argued that
because of the one-component nature of the dimerization order
parameter for a spin-Peierls phase transition, asymptotically
the transition must be in the 3D Ising universality class.
They argued further, that because of the coupling to the
elastic strains, the precritical behavior should be
tricritical-like.  Similar conclusions, albeit based on
different physical reasoning, were arrived at later by Werner
and Gros \cite{Werner}. 
Proponents of 3D XY behavior typically argue that the copper
and oxygen displacements are independent thence yielding a two-component
order parameter \cite{Plumer}. 
Implicitly, Harris {\it et al.} \cite{Harris} assume that all of the atomic
displacements accompanying the spin-Peierls transition are linearly
coupled thence reducing the system to a one-component order parameter.
The 3D critical behavior models seem to be supported
by measurements of the order parameter \cite{Harris,Lum,Lorenzo} 
which for reduced
temperatures
$\sim 2\times 10^{-3}<1-T/T_{SP}\lesssim\ 5\times
10^{-2}$ exhibits power law behavior
$(1-T/T_{SP})^\beta$ with $\beta = 0.33\pm 0.02$, in good agreement
with both 3D Ising and XY values of $\beta = 0.325$ and 0.345
respectively \cite{Le}. The heat capacity data are equally well described
by a 3D critical behavior model (Ising or XY) and by a mean field
model with Gaussian fluctuations \cite{Lasj}.

In this paper we present an alternative model for CuGeO$_3$,
namely a Landau-Ginzburg model incorporating a tricritical to
mean field crossover. As we shall show, this model describes all
available data very well with few adjustable parameters. The format
of this paper is as follows. In Section II we introduce the model
including its genesis in studies of critical phenomena in thermotropic
liquid crystals systems. Section III presents an analysis of the 
available data for CuGeO$_3$ using this model. Finally, in Section IV
we give a summary, our conclusions and suggestions for future experiments.

\section{The Model}

The conundrum described above for CuGeO$_3$
is reminiscent of a similar divergence of views
which occurred in the interpretation of experiments on smectic A -
smectic C phase transitions in thermotropic liquid crystal systems
\cite{Gal}-\cite{Birgeneau}.
In particular, in that case, measurements of the tilt 
order parameter \cite{Gal,Birgeneau}
typically reveal power law behavior
$\phi \sim (1-T/T_{AC})^\beta$ over the temperature range
$5\times 10^{-5} < (1-T/T_{AC}) < 5\times 10^{-3}$  with
$\beta = 0.36 \pm 0.02$.  However, this divergence of views was resolved by
Huang and Viner \cite{Huang}
and Birgeneau {\it et al.} \cite{Birgeneau} who showed
that all of the data including the heat capacity, order parameter, and
tilt susceptibility, were consistent with the predictions of a simple
Landau model with an anomalously large 6th order term.
Clearly, it is of interest to carry out a similar analysis for
the available data for the spin-Peierls transition in
CuGeO$_3$.

For the Landau-Ginzburg model the free energy is given by
\begin{equation}
F = a \tau \phi^2 + b \phi^4 + c \phi^6 .... + {1\over
2m_\alpha} |\nabla_\alpha \phi|^2 \end{equation}
where $\tau = T/T_c -1$.

With $\tau_0= b^2/ac$, standard calculations yield for the order
parameter, $\phi$, the specific heat, $C$, the susceptibility, $\chi$, and
the correlation length, $\xi_\alpha$,:

\begin{eqnarray}  \phi &=& (b/3c)^{1/2} [(1-3\tau/\tau_0)^{1/2} -1]^{1/2}
\hspace{0.50in}
\tau < 0\\ C & = & \left\{ \begin{array}{lr} 0  & \tau > 0 \\
(a^2 T/2bT_c^2) (1-3 \tau/\tau_0)^{-1/2} & \hspace{0.5in} \tau < 0
\end{array} \right.\\
\chi &=& 1/2a\tau \hspace{2.20in} \tau > 0 \\
\xi_\alpha &=& (2am_\alpha\tau)^{-1/2} \hspace{1.75in} \tau > 0
\end{eqnarray}
with similar expressions for $\tau < 0$ for $\chi$ and $\xi$.
Eq. (2) and (3) are conveniently rewritten in the form
\begin{eqnarray} \begin{array}{llll} \phi &=& \phi_0 
\begin{array}{ll} \left[ \left( 1
+ 3
\frac{T_{SP} - T}{T_{SP}-T_{CR}} \right)^{1/2}-1 \right]^{1/2} &
\hspace{0.25in} \tau < 0 \end{array} \end{array}  \\
\begin{array}{llll} C & = &\left\{ \begin{array}{ll}  0 & \hspace{0.5in}
\tau > 0 \\
C\_T\left( 1 + 3 \frac{T_{SP} - T}{T_{SP}-T_{CR}}
\right)^{-1/2} & \hspace{0.5in} \tau < 0 \end{array}\right.\end{array}
\end{eqnarray}
where $T_{CR}$ is the crossover temperature from tricritical to mean
field behavior.  We note that in the above expressions the
exponents are fixed and only the amplitudes and the
two temperatures, $T_{SP}$ and
$T_{CR}$, are variable.  A log-log plot of Eq. (6) reveals that for the
order parameter $\phi$ the effective exponent $\beta$ crosses over
gradually from
$1\over 4$ to $1\over 2$ as $T$ varies from less than to greater than
$T_{CR}$. In the smectic A - smectic
C case the measurements span $T_{CR}$ and accordingly intermediate
exponents, $\beta \simeq 0.36$, are found even though the actual
transition is mean-field-like for temperatures in the immediate
vicinity of $T_{AC}$ \cite{Birgeneau}.

\section{Analysis}

We now apply this tricritical--mean field crossover model to CuGeO$_3$.
The first test is $T_{SP}$ itself or, more precisely, the ratio of the
spin gap, $\Delta$, to $T_{SP}$.  In the mean field theory of 
Pytte \cite{Pytte}, the
spin-Peierls transition is BCS-like so that in the weak coupling limit
$2\Delta / T_{SP} = 3.5$.  In the charge transfer salts TTF -
CuBDT \cite{Bray,Mon},
TTF - AuBDT \cite{Jac}, 
MEM - (TCNQ)$_2$ \cite{Hu}, and SBTTF - TCNQCl$_2$ \cite{Jac1} this ratio is
found to
be 3.5, 3.7, 3.1 and  $\leq$ 3.5 respectively, in good agreement with the BCS
value.  Critical fluctuations, either Ising or XY in character, would act
to increase this ratio.  For CuGeO$_3$, $\Delta = 24.5$K and $T_{SP}
\simeq 14$K implying
$2\Delta/T_{SP} = 3.5$, consistent with a BCS mean field theory
description \cite{Hase,Nishi}. At the minimum, this value for $2\Delta/T_{SP}$
argues against any quantitatively important effect of critical
fluctuations on $T_{SP}$ in CuGeO$_3$.

The behavior of the order parameter in CuGeO$_3$ is of
particular importance since this observable appears to provide
the strongest evidence for true critical rather than mean field or
tricritical behavior.  A number of groups have reported
measurements of the temperature dependence of the order parameter
in CuGeO$_3$ \cite{Harris,Lum,Lorenzo}.  
The measured phase transition 
temperature
$T_{SP}$  varies between 13.3K and 14.6K in different
samples.  Nevertheless, near-universal behavior is observed for
the order parameter provided that it is plotted as a function
of the reduced temperature $T/T_{SP}$.  As noted above, fits
of the order parameter $\phi (T/T_{SP})$ for $1-T/T_{SP} <
0.05$ to a single power law $\phi\sim (1-T/T_{SP})^\beta$ all yield
values of $\beta = 0.33 \pm 0.02$.  As discussed by Gaulin
and co-workers \cite{Lum}, inclusion of a correction-to-scaling
multiplicative factor $(1+B|\tau |^\delta)$ in the expression for
$\phi$ both improves the goodness of fit and, not surprisingly,
extends the range of validity of the fit.

\begin{figure}
\centerline{\epsfxsize=2.9in\epsfbox{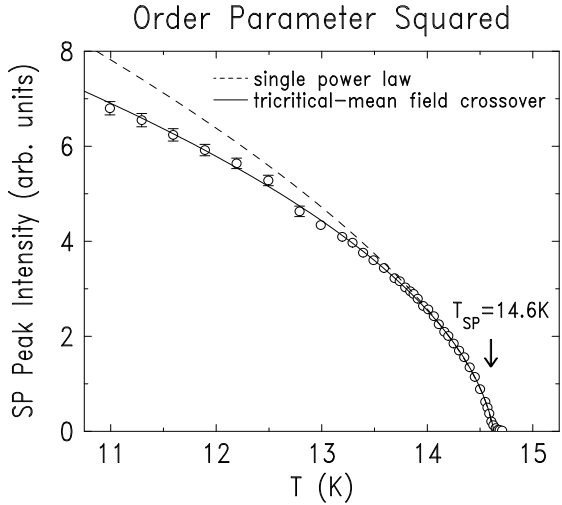}}
\vskip 5mm
\caption{ (3/2, 1, 3/2) superlattice peak intensity measured
with synchrotron x-ray diffraction techniques.  The peak intensity
is proportional to the order parameter squared, $\phi^2$.  The
dashed line is the result of a fit of the data for
$\tau = 1-{{T}\over{T_{SP}}} < 0.04$ to a single power law $\phi^2 \sim
\phi^2_0 (1-{{T}\over{T_{SP}}})^{2\beta}$ with $\beta = 0.314
\pm 0.01$.  The solid line is the result of a fit to Eq. (6) with
$\tau_{CR} = 0.006$.}
\label{fig1}
\end{figure}

We show in Fig. 1 our own measurements of the order
parameter squared in a sample of CuGeO$_3$ with $T_{SP} =
14.6$K.  These data are consistent with those measured by both
ourselves and other groups in a variety of samples
\cite{Harris,Lum,Lorenzo}.  Fits to a
single power law for $\tau < 0.04$ yield $\beta = 0.314 \pm 0.01$.
However, as noted by Harris {\it et al.} \cite{Harris} and as may be
seen in Fig. 1, the data fall significantly below the power
law curve for $\tau > 0.04$.  We show, in addition, in Fig. 1
the results of a fit to the tricritical to mean field
crossover form, Eq. (6). This fit has only three adjustable
parameters, $\phi^2_0$, $T_{CR}$ and $T_{SP}$. This is the same
number of parameters as those in the single power law fits
discussed above and two less than the number of adjustable
parameters in fits to a power law with corrections-to-scaling with both
$B$ and $\delta$ varied. 
(We note that Lumsden {\it et al.} \cite{Lum} fix $\delta =1/2$ whereas
Lorenzo {\it et al.} allow $\delta$ to vary; the latter group find
an optimum fit for $\delta\simeq 1$).
It is evident that Eq. (6) describes the order
parameter data extremely well over the complete range of temperatures.
The fit yields $\tau_{CR} = 1-T_{CR}/T_{SP} = 0.006 \pm 0.001$ implying
that the crossover from tricritical to mean field behavior occurs at a
quite small reduced temperature.

We now discuss the energy gap $\Delta$.  Using a simple
scaling ansatz, Cross and Fisher \cite{Cross} argue that $\Delta \sim
\phi ^{2/3}$.  We show in Fig. 2 the data of Lorenzo {\it et
al.} \cite{Lorenzo} for the magnetic energy gap for $T < T_{SP}$ in a sample of
CuGeO$_3$ with
$T_{SP} = 14.4$K.  In part because of the apparent jump
of $\Delta(T)$ at $T_{SP}$, Lorenzo {\it et al.} \cite{Lorenzo} interpret
these data as indicating a 2D XY Kosterlitz-Thouless
transition \cite{KT}.  In fact, these data are readily explained using the
model of Cross and Fisher \cite{Cross} together with the tricritical-mean field
crossover form for $\phi$, Eq. (6).  In this case we hold $T_{SP}$ fixed
at $T_{SP} = 14.4$K and set $\tau_{CR} = 0.006$ as determined above so
that there is only one adjustable  parameter, the overall amplitude
$\Delta(0)$.  The result so-obtained is shown in Fig. 2.  It
is evident that the tricritical-mean field model with
$\Delta(T) \sim \phi^{2/3}$ describes the measured gap energy
$\Delta(T)$ extremely well over a wide range of temperatures with
only one adjustable parameter.  Indeed, this is by far the best
test to-date of the Cross-Fisher model.  We should note that this model cannot
explain the inferred pseudogap above T$_{SP}$ \cite{Lorenzo}. However, the
``pseudogap'' is
deduced using a heuristic line-shape analysis which lacks a firm
theoretical basis.

\begin{figure}
\centerline{\epsfxsize=2.7in\epsfbox{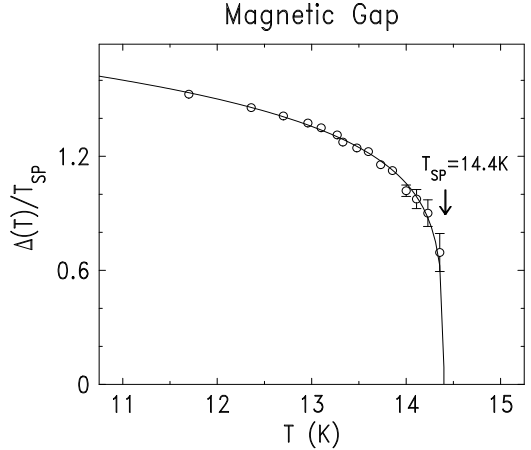}}
\vskip 5mm
\caption{Magnetic energy gap in CuGeO$_3$. These data are from
Ref. 17.  The
solid line is the result of a fit to the form $\Delta(T) = \Delta(0)
\phi^{2/3}$ where $\phi$ is given by Eq. (6) with $\tau_{CR}$ held
fixed at 0.006.}
\label{fig2}
\end{figure}

The specific heat in CuGeO$_3$ has proven to be the most difficult
thermodynamic quantity to interpret unambiguously \cite{Lasj}.  This is, in
part, because of the extreme sensitivity of the specific heat near
$T_{SP}$ to sample inhomogeneities and, in part, because of the
inevitable large number of adjustable parameters required to
describe the critical specific heat in any physically relevant
model.  Fig. 3 shows high resolution magnetic specific heat $(C_M)$ data
for a sample of CuGeO$_3$ with $T_{SP} = 14.24$K from Lasjaunias and
coworkers \cite{Lasj}.  Hegman {\it et al.} \cite{Lasj}
have carried out an extensive
analysis of these data using both a mean field  ``BCS plus Gaussian
fluctuation'' model and a critical behavior model.  They find that
both models describe $C_M$ quite well in the immediate vicinity of
$T_{SP}$, albeit at the cost of a rather large number of
adjustable parameters.  The critical behavior model fits give a value for
the specific heat exponent, $\alpha$, near 0.  On the other
hand, the Gaussian fluctuation analysis implies that the true
critical behavior is confined to the region $|\tau| < 0.0006$.

\begin{figure}
\centerline{\epsfxsize=2.7in\epsfbox{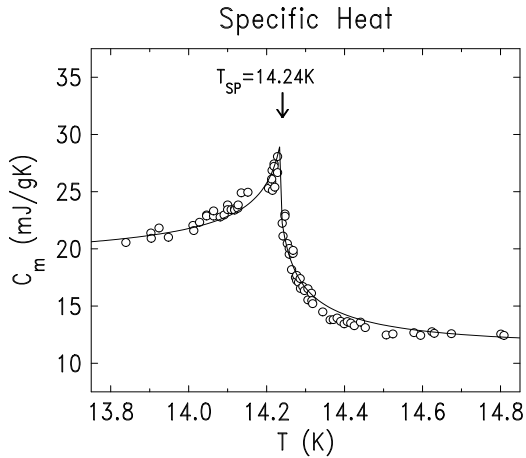}}
\vskip 5mm
\caption{Magnetic specific heat in CuGeO$_3$. These data are from
Ref. 20.
The solid line is the result of a fit to Eq. (8) with $\tau_{CR}$ held fixed at
0.006.}
\label{fig3}
\end{figure}

Given the uncertainties connected with the fits described above, the best
one can hope for is to determine whether or not the
tricritical-mean field crossover model is consistent with the
experimental results for $C_M$ shown in Fig. 3.  First, it is evident that
Eq. (7) will be inadequate since one must, at the minimum, include
Gaussian fluctuations above $T_{SP}$.  We therefore 
include the fluctuations above T$_{SP}$ in the simplest way possible
by replacing Eq. (7) by
\begin{equation}\begin{array}{lrl} C_M &=& \left \{
\begin{array} {lr}C^+_MT
\left(1 + 3 {{T-T_{SP}}\over{T_{SP}-T_{CR}}}\right)^{-1/2} +\gamma T
&\hspace{0.30in} \tau > 0 \\
C^-_MT\left( 1 + 3 {{T_{SP}-T}\over{T_{SP}- T_{CR}}}\right)^{-1/2}
+ B\_ &\hspace{0.30in} \tau < 0 \end{array}\right.\end{array}\end{equation}
where $\gamma T$ is the regular linear term for a 1D Heisenberg
antiferromagnet and $B\_$ is the background term below
$T_{SP}$. 
The background $B_-$ should, in general, be temperature dependent;
however, given the narrow range of temperatures we consider, a
constant background is adequate.
Eq. (8) is closely similar to the BCS plus Gaussian
fluctuation model considered by Hegman {\it et al.} \cite{Lasj}
since the Gaussian
fluctuations give rise to a
$|\tau|^{-1/2}$ contribution to $C_M$ both above and
below $T_{SP}$.  The solid lines in Fig. 3 correspond to fits to
Eq. (8) with $\tau_{CR}$ fixed at 0.006 and $C^+_M$,
$C^-_M$,$\gamma$, $B\_$ and $T_{SP}$ varied.  Clearly Eq. (8)
describes $C_M$ quite well; indeed the fit appears to be better
than those for either of the models tested by Hegman {\it et al.} \cite{Lasj}.
The fit shown in Fig. 3 gives $C^+_M/C^-_M = 1.1 \pm 0.13$; 
this ratio is expected to be non-universal so it cannot be simply interpreted.
We conclude, therefore, that the tricritical-mean field crossover
model describes $C_M$ well although not uniquely so.

Finally, we discuss the correlation length and the staggered
susceptibility.  Pouget {\it et al.} \cite{Pouget}
have found that the correlation
length over a wide temperature range follows the behavior $\xi
\sim (T/T_{SP}-1)^{-1/2}$, consistent with mean field theory;
however, the number of data points in their experiment near $T_{SP}$ is
sufficiently small that their results do not meaningfully
differentiate between various theoretical models.  
Harris {\it et al.} \cite{Harris}
have reported a high resolution synchrotron x-ray study of the critical
fluctuations above $T_{SP}$ in CuGeO$_3$.  They find pretransitional lattice
fluctuations within 1K above $T_{SP}$ whose length scale is about an order of
magnitude longer than those characterizing the bulk thermal fluctuations.  The
line-shape of the large length scale fluctuations is consistent with a
Lorentzian-squared form.  The measured critical exponents are
$\nu = 0.56 \pm 0.09$ and $\bar\gamma = 2.0 \pm 0.3$ where
$\bar \gamma$ is the exponent characterizing the divergence of the
disconnected staggered susceptibility \cite{Mukamel}. 
The mean field predictions
for these exponents are $\nu = 1/2$ and $\bar\gamma = 2\gamma
= 2$ whereas for 3D Ising (XY) critical behavior one expects
$\nu = 0.63 \ (0.67)$ and $\bar\gamma = 2.5 (2.64)$.  Thus
the Harris {\it et al.} \cite{Harris}
data favor the tricritical-mean field model
but 3D Ising or XY critical models are not excluded.
Precise measurements of the bulk staggered susceptibility using
neutrons should yield accurate values for $\nu$ and $\bar\gamma$ and
this, in turn, would definitively choose between the models.

\section{Discussion}

In summary, each of the order parameter, magnetic energy gap,
specific heat, correlation length and disconnected staggered
susceptibility are well-described by a simple Landau-Ginzburg
model exhibiting a tricritical-mean field crossover near
$T_{SP}$.  Further, the ratio of the energy gap to $T_{SP}$ is
consistent with the value for a BCS mean-field
transition.  We conclude, therefore, that CuGeO$_3$, in common with
the organic change transfer salts, exhibits a mean field
spin-Peierls transition for reduced temperatures $|\tau|> 0.001$.

The principal remaining issue is the microscopic origin of the
tricritical behavior.  Harris {\it et al.} \cite{Harris} argue that this is
caused by a diminution in the effective fourth order term in Eq.
(1), $b \phi^4$, because of coupling to the macroscopic strain.
It also seems possible that competing nearest and next-nearest
neighbor exchange interactions along the chain could generate the
tricritical instability \cite{Werner,Castilla}. Specifically, Castilla {\it et
al.} \cite{Castilla}
argue that the ratio of the next nearest neighbor to nearest
neighbor exchange interaction along the chain is close to the critical
value for spontaneous formation of a magnetic gap independent of coupling
to the lattice.  Heuristically, it seems that this could generate
tricritical behavior in the phase diagram.  Another possible
source of tricritical behavior is competition between the N\'{e}el
state and the spin-Peierls state, that is, competition between
the coupling of the $S=1/2 \ $chain to the lattice and the interchain
exchange coupling.  Clearly, a multidimensional theoretical analysis of the
spin-Peierls phase diagram including magnetostriction, competing intrachain
exchange
interactions together with the interchain magnetic and elastic coupling is
required.

Of course, the mean field behavior itself in all of these spin-Peierls
systems is not yet well understood. In TTF-CuBDT there is evidence for a
soft phonon at very high temperatures \cite{Mon} and Cross and Fisher
\cite{Cross} speculate that the precursive soft mode accounts for the
large length scale underlying the mean field behavior. In CuGeO$_3$,
no soft phonon at all has yet been seen. Thus, the microscopic origin
of the large length scale in CuGeO$_3$ remains to be elucidated. 

Finally, it would be very interesting to see if the putative nearby
tricritical point could be accessed by changing some variable such as
pressure, uniaxial stress or doping.  Masuda {\it et al.} \cite{Masuda} have
shown that replacement of Cu by Mg both depresses $T_{SP}$ and appears to
drive the spin-Peierls transition first order.  The concomitant
tricritical point could well account for the observed tricritical-mean
field crossover in pure CuGeO$_3$.  We note, however, that the actual
physics of magnetic dilution in CuGeO$_3$ is quite complex since dilution
introduces frustration of the interchain elastic interaction \cite{Wang}. 
Replacement of Cu$^{2+}$ by Cd$^{2+}$ (Ref. \cite{Cd}) or Ge$^{4+}$ by
Ga$^{4+}$ (ref. \cite{Ga}) both lead to mean field behavior over quite
wide temperature ranges; that is, doping with these ions moves CuGeO$_3$
away from the tricritical point into the pure mean field regime. Again,
further research, both experimental and theoretical, is required to
elucidate these effects more completely.

\section{Acknowledgements}

We thank N. Hegman and J. C. Lasjaunias for comments on the manuscript,
for sending us their
results for $C_M$ in tabular form and for a preprint of Ref. \cite{Lasj}.
We are grateful to C. W. Garland and N. Goldenfeld for insightful
communications on the interpretation of the critical heat capacity.
We thank B. D. Gaulin for valuable comments on this paper and for
drawing our attention to Ref. \cite{Cd}. This
work was supported by the NSF under Grant No. DMR97-04532 and by the
MRSEC Program of the NSF under award No. DMR98-08941.


\end{document}